\documentstyle[11pt,fleqn]{article}  
\input{psfig.tex}                
\title{Possible Orientation Effects to the Phase Diagram of Fundamental Composite Particles}
\author{Kwang-Hua  W. Chu} 
\date{P.O. Box 39, Transfer Centre, Road XiHong, Wulumuqi 830000, PR China}
\begin{document}      
\maketitle              
\begin{abstract}           
We obtain possibly valuable information about the phase diagram
of rather dense composite particles at high fermion as well as boson number density
but low temperature, which is not
accessible to relativistic heavy ion collision experiments.  
Our calculations
could qualitatively measure the number of
thermodynamic degrees of freedom  at the time  at which
the matter (produced in heavy ion collisions which might be a deconfined
quark-gluon plasma) comes into approximate
local thermal equilibrium and begins to behave like a hydrodynamic fluid.
Possible
observational signatures associated with the theoretically proposed states
of matter inside compact
stars are discussed as well.
\end{abstract}
\doublerulesep=6mm    
\baselineskip=6mm
\bibliographystyle{plain}
\section{Introduction}
The only place in the universe where we expect sufficiently high densities
and low temperatures is compact stars, also known as 'neutron stars', since
it is often assumed that they are made primarily of neutrons (for a recent
review, see [1]). A compact star is produced in a supernova. As the outer
layers of the star are blown off into space, the core collapses into a very dense
object.
Neutron stars are among the most fascinating bodies in our Universe. They contain
over a solar mass of matter within a radius of $\sim$10 km at densities of the order
of 10$^{15}$ g/cc. They probe the properties of cold matter at extremely high densities
and have proven to be fantastic test bodies for theories of general relativity.
In a broader perspective, neutron stars and heavy-ion collisions provide access
to the phase diagram of matter at extreme densities and temperatures, which is
basic for understanding the very early Universe and several other astrophysical
phenomena. \newline
Astrophysicists distinguish between three different kinds of compact stars. These are white
dwarfs, neutron stars, and black holes. The former contain matter in one of the densest forms found
in the Universe which, together with the unprecedented progress in observational astronomy, makes
such stars superb astrophysical laboratories for a broad range of most striking physical phenomena.
These range from nuclear processes on the stellar surface to processes in electron degenerate matter
at subnuclear densities to boson condensates and the existence of new states of baryonic matter〞
such as color superconducting quark matter at supernuclear densities. More than that, according
to the strange matter hypothesis strange quark matter could be more stable than nuclear matter, in
which case neutron stars should be largely composed of pure quark matter possibly enveloped in
thin nuclear crusts. 
Neutron stars and white dwarfs are in hydrostatic
equilibrium, so
at each point inside the star gravity is balanced by the degenerate
particle pressure,
as described mathematically by the Tolman-Oppenheimer-Volkoff equation [2].\newline
It is often stressed that there has never been a more exciting time
in the overlapping
areas of nuclear physics, particle physics, and relativistic astrophysics
than today [3]. This
comes at a time where new orbiting observatories such as the Hubble Space Telescope,
Rossi X-ray Timing Explorer (RXTE), Chandra X-ray satellite, and X-ray Multi Mirror
Mission (XMM) have extended our vision tremendously, allowing us to see vistas with an
unprecedented clarity and angular resolution that previously were only imagined, enabling
astrophysicists for the first time ever to perform detailed studies of large samples of galactic
and extragalactic objects. On the Earth, radio telescopes (e.g., Arecibo, Green Bank,
Parkes, VLA) and instruments using adaptive optics and other revolutionary techniques
have exceeded previous expectations of what can be accomplished from the ground. The
gravitational wave detectors LIGO, LISA, VIRGO, and Geo-600 are opening up a window
for the detection of gravitational waves emitted from compact stellar objects such as
neutron stars and black holes.\newline
Neutron stars are dense, neutron-packed remnants of massive stars that blew apart in
supernova explosions [1]. They are typically about twenty kilometers across and spin rapidly,
often making several hundred rotations per second. Many neutron stars form radio pulsars,
emitting radio waves that appear from the Earth to pulse on and off like a lighthouse beacon
as the star rotates at very high speeds. Neutron stars in X-ray binaries accrete material
from a companion star and flare to life with a burst of X-rays. Measurements of radio
pulsars and neutron stars in X-ray binaries comprise most of the neutron star observations.
Improved data on isolated neutron stars (e.g. RX J1856.5-3754, PSR 0205+6449) are now
becoming available, and future investigations at gravitational wave observatories such as
LIGO and VIRGO will focus on neutron stars as major potential sources of gravitational
waves. Depending on star mass and rotational frequency, gravity compresses the matter
in the core regions of pulsars up to more than ten times the density of ordinary atomic
nuclei, thus providing a high pressure environment in which numerous subatomic particle
processes compete with each other.\newline
The most spectacular ones stretch from the generation
of hyperons and baryon resonances to quark deconfinement to
the formation of boson condensates. There are theoretical
suggestions of even more exotic processes inside neutron stars, such as the formation of
absolutely stable strange quark matter, a configuration of matter more stable than
the most stable atomic nucleus, $^{62}$Ni. 
Instead
these objects should be named nucleon stars, since relatively isospin symmetric nuclear
matter-in equilibrium with condensed $K^-$ mesons-may prevail in their interiors [3],
hyperon stars if hyperons ($\Sigma$,$\Lambda$,$\Xi$, possibly in
equilibrium with the $\Delta$ resonance) become
populated in addition to the nucleons, quark hybrid stars if the highly compressed
matter in the centers of neutron stars were to transform into u, d, s quark matter,
or strange stars if strange quark matter were to be more stable than
nuclear matter. 
Of course, at present one does not know from experiment at what density
the expected phase transition to quark matter occurs. Neither do lattice Quantum
ChromoDynamical (QCD) simulations provide a conclusive guide yet. From simple
geometrical considerations it follows that, for a characteristic nucleon radius of $r_N
\sim 1$fm, nuclei begin to touch each other at densities of $\sim (4 \pi r^3_N/3))^{-1}
 \approx 0.24$ fm$^3$, which
is less than twice the baryon number density of ordinary nuclear matter, $\rho_0 = 0.16$ fm$^3$ (energy density  $\epsilon_0 = 140$ MeV/fm$^3$). Depending on the rotational frequency and stellar
mass, such densities are easily surpassed in the cores of neutron stars so gravity may
have broken up the neutrons (n) and protons (p) in the centers of neutron stars into their
constituents. 
The phase diagram of quark matter, expected to be in a color
superconducting phase, is
very complex [3,4]. At asymptotic densities the ground state of QCD with a vanishing
strange quark mass is the color-flavor locked (CFL) phase. This phase is electrically charge neutral without any need for electrons for a significant range of chemical potentials and
strange quark masses.\newline
An intriguing aspect of quantum mechanics is that even at absolute
zero temperature quantum fluctuations prevail in a system, whereas
all thermal fluctuations are frozen out. These quantum
fluctuations are able to induce a macroscopic phase transition in
the ground state of a many-body system, when the ratio of two
competing terms in the underlying Hamiltonian is varied across a
critical value [5]. The instability of the quantum critical
behavior with respect to (w.r.t.) the disorder can be interpreted
as a signal for phenomena of localization. 
One example
is the strongly-interacting electron (non-Fermi) liquid with
different strengths of disorder. Interesting
issues are the quantum phase transition (QPT) and quantum critical
point (QCP). \newline
Quite recently McLaughlin {\it et al.} [6]
searched for radio sources that vary on much
shorter timescales. They found eleven objects characterized by
single, dispersed bursts having durations between 2 and 30 ms.
The average time intervals between bursts range from 4 min to 3 h
with radio emission typically detectable for $<1$ s per day. From an
analysis of the burst arrival times, they have identified periodicities
in the range 0.4-7 s for ten of the eleven sources, suggesting origins
in rotating neutron stars. Meanwhile as all pulsars from which
giant pulses have been detected appear to have high values of
magnetic field strength at their light cylinder radii [7]. While the Crab
pulsar has a magnetic field strength at the light cylinder of
9.3 $\times 10^5$ G [6], this value ranges from only 3 to 30G for these sources,
suggesting that the bursts originate from a different emission
mechanism. They therefore concluded that these sources represent a previously
unknown population of bursting neutron stars, which they call
rotating radio transients (RRATs).  These interesting new observations make us
to investigate the rotation (e.g., it is still controversial how much
angular momentum the iron cores
have before the onset of the gravitational-collapse) [1,6] as well as
Pauli-blocking effects [8] in compact stars
considering the phase diagram [4].
\newline
%
In present approach the
{Ue}hling-Uhlenbeck collision term [9-10] which could describe the
collision of a gas of dilute hard-sphere Fermi- or Bose-particles
by tuning a parameter $\gamma$ : a Pauli-blocking factor (or
$\gamma f$ with $f$ being a normalized (continuous) distribution
function giving the number of particles per cell) is adopted
together with a  free-orientation $\theta$ (which is
related to the relative direction of scattering of particles
w.r.t. to the normal of the propagating plane-wave front) into the
quantum discrete kinetic model [10] which can be used to obtain
dispersion relations of plane (sound) waves propagating in different-statistic
gases (of particles). We then study the  critical behavior based on the
acoustical analog [12] which has been verified before. The
possible phase diagram  (as the temperature is
changed) related to the effective number of thermodynamic
degrees of freedom (which is useful in QCD thermodynamics, it is one
observable
whose value changes by more than an order of magnitude
across the crossover transition from hadron gas to
quark-gluon plasma (QGP)[11]) we obtained resemble qualitatively those
proposed before.
\section{Theoretical Formulations}
Within this presentation,
we shall assume that the gas is composed of identical hard-sphere
particles of the same mass {\cite{U:U}}.
(Note that most of the similar  formulations could be traced in Chu
\cite{U:U} or [12-13]) The
velocities of these particles are restricted to, e.g., : ${\bf
u}_1, {\bf u}_2, \cdots, {\bf u}_p$, $p$ is a finite positive
integer. The discrete number densities of particles are denoted by
$N_i ({\bf x},t)$ associated with the velocities ${\bf u}_i$ at
point ${\bf x}$ and time $t$. If only nonlinear binary collisions
and the evolution of $N_i$ are considered, we have
\begin{equation}
 \frac{\partial N_i}{\partial t}+ {\bf u}_i \cdot \nabla N_i
 = F_i \equiv \frac{1}{2}\sum_{j,k,l} (A^{ij}_{kl} N_k N_l - A_{ij}^{kl}
 N_i N_j),
\hspace*{3mm} i \in\Lambda =\{1,\cdots,p\},
\end{equation}
where $(i,j)$ and $(k,l)$ ($i\not=j$ or $k\not=l$) are admissible
sets of collisions
\cite{U:U}.
Here, the summation is taken over all $j,k,l \in \Lambda$, where
$A_{kl}^{ij}$ are nonnegative constants satisfying
\cite{U:U}
 $ A_{kl}^{ji}=A_{kl}^{ij}=A_{lk}^{ij}$, 
 $ A_{kl}^{ij} ({\bf u}_i +{\bf u}_j -{\bf u}_k -{\bf u}_l )=0$,
 and $A_{kl}^{ij}=A_{ij}^{kl}$. 
The conditions defined for the discrete velocities above require
that there are elastic, binary collisions, such that momentum and
energy are preserved, i.e.,
 ${\bf u}_i +{\bf u}_j = {\bf u}_k +{\bf u}_l$,
 $|{\bf u}_i|^2 +|{\bf u}_j|^2 = |{\bf u}_k|^2 +|{\bf u}_l|^2$,
are possible for $1\le i,j,k,l\le p$.  
We note that, the summation of $N_i$ ($\sum_i N_i$) : the total
discrete number density here is related to the macroscopic density
: $\rho \,(= m_p \sum_i N_i)$, where $m_p$ is the mass of the
particle \cite{U:U}.
\newline
Together with the introducing of the {U}ehling-Uhlenbeck
collision term \cite{U:U} :
 $F_i$ $=\sum_{j,k,l} A^{ij}_{kl} \,[ N_k N_l$ $(1+\gamma N_i)(1+\gamma N_j)$ $-
 N_i N_j (1+\gamma N_k)(1+\gamma N_l)]$,
into equation (1), for $\gamma <0$ (normally, $\gamma=-1$), we can
then obtain a quantum discrete kinetic equation for a gas of
Fermi-particles; while for $\gamma
> 0$ (normally, $\gamma=1$) we obtain one for a gas of Bose-particles,
and for $\gamma =0$ we recover the equation (1).  \newline
Considering binary  collisions only, from equation above, the
model of quantum discrete kinetic equation for Fermi or Bose gases
proposed before is then a system of $2n(=p)$ semilinear
partial differential equations of the hyperbolic type :
\begin{displaymath}
 \frac{\partial}{\partial t}N_i +{\bf v}_i \cdot\frac{\partial}{\partial
 {\bf x}} N_i =\frac{c S}{n} \sum_{j=1}^{2n} N_j N_{j+n}(1+\gamma N_{j+1})
 (1+\gamma N_{j+n+1})-
\end{displaymath}
\begin{equation}
 \hspace*{18mm} 2 c S N_i  N_{i+n} (1+\gamma N_{i+1})(1+\gamma
 N_{i+n+1}),\hspace*{24mm} i=1,\cdots, 2 n,
\end{equation}
where $N_i=N_{i+2n}$ are unknown functions, and ${\bf v}_i$ =$ c
(\cos[\theta+(i-1) \pi/n], \sin[\theta+(i-1)\pi/n])$; $c$ is a
reference velocity modulus and the same order of magnitude as that
 used in Ref. 10 ($c$, the sound speed in the absence of scatters),
$\theta$ is the orientation starting from the positive $x-$axis to
the $u_1$ direction,
$S$ is an effective collision cross-section for the collision
system.
\newline Since passage of the plane (sound wave) will cause a small departure
from an equilibrium state and result in energy loss owing to
internal friction and heat conduction, we linearize above
equations around a uniform equilibrium state (particles' number
density : $N_0$) by setting $N_i (t,x)$ =$N_0$ $(1+P_i (t,x))$,
where $P_i$ is a small perturbation. 
After some similar manipulations  (please refer to Chu in [12-13]), with
$B=\gamma N_0 <0$, which gives or defines the
(proportional) contribution from the Fermi gases (if $\gamma < 0$,
e.g., $\gamma=-1$) or the Bose gases ($B>0$, if $\gamma > 0$,
e.g., $\gamma=1$), we then have
\begin{equation}
 [\frac{\partial^2 }{\partial t^2} +c^2
 \cos^2[\theta+\frac{(m-1)\pi}{n}]
 \frac{\partial^2 }{\partial x^2} +4 c S N_0 (1+B) \frac{\partial
 }{\partial t}] D_m= \frac{4 c S N_0 (1+B)}{n} \sum_{k=1}^{n} \frac{\partial
 }{\partial t} D_k  ,
\end{equation}
where $D_m =(P_m +P_{m+n})/2$, $m=1,\cdots,n$, since $D_1 =D_m$
for $1=m$ (mod $2 n)$. \newline 
We are ready to look for the solutions in the form of plane wave
$D_m$= $a_m$ exp $i (k x- \omega t)$, $(m=1,\cdots,n)$, with
$\omega$=$\omega(k)$. This is related to the dispersion relations
of 1D (forced) plane wave propagation in Fermi or Bose gases. So we have
\begin{equation}
 (1+i h (1+B)-2 \lambda^2 cos^2 [\theta+\frac{(m-1)\pi}{n}]) a_m -\frac{i h (1+B)}{n}
 \sum_{k=1}^n a_k =0  , \hspace*{6mm} m=1,\cdots,n,
\end{equation}
where
\begin{displaymath}
\lambda=k c/(\sqrt{2}\omega),  \hspace*{18mm} h=4 c S N_0 /\omega
\hspace*{6mm} \propto \hspace*{2mm} 1/K_n,
\end{displaymath}
where $h$ is the rarefaction parameter of the gas; $K_n$ is the
Knudsen number which is defined as the ratio of the mean free path
of gases to the wave length of the plane (sound) wave.
\newline
\section{Numerical Results and Discussions}
We firstly introduce the concept of acoustical
analog [12] in brief.
In a mesoscopic system, where the sample size is smaller than the
mean free path for an elastic scattering, it is satisfactory for a
one-electron model to solve the time-independent Schr\"{o}dinger
equation :
 $-({\hbar^2}/{2m}) \nabla^2 \psi + V' (\vec{r}) \psi = E \psi$
or (after dividing by $-\hbar^2/2m$)
 $\nabla^2 \psi + [q^2 - V (\vec{r})] \psi = 0$,
where $q$ is an (energy) eigenvalue parameter, which for the
quantum-mechanic system is $\sqrt{2mE/\hbar^2}$. Meanwhile, the
equation for classical (scalar) waves is
 $\nabla^2 \psi - ({\partial^2 \psi}/{c^2 \,\partial t^2})
 =0$
or (after applying a Fourier transform in time and contriving a
system where $c$ (the wave speed) varies with position $\vec{r}$)
 $\nabla^2 \psi + [q^2 - V (\vec{r})] \psi = 0$,
here, the eigenvalue parameter $q$ is $\omega/c_0$, where $\omega$
is a natural frequency and $c_0$ is a reference wave speed.
Comparing the time dependencies one gets the quantum and classical
relation $E= \hbar \omega$. The localized state could thus
be determined via $E$ or the
rarefaction parameter ($h$) which is related to the ratio of the collision frequency
and the wave frequency [12].\newline
The complex spectra ($\lambda=\lambda_r +$ i
$\lambda_i$; the real part $\lambda_r = k_r c/(\sqrt{2}\omega)$: sound
dispersion, a relative measure of the sound or phase speed;
the imaginary part $\lambda_i = k_i c/(\sqrt{2}\omega)$ : sound attenuation or
absorption) could be obtained from the complex polynomial equation above. Here,
the Pauli-blocking parameter ($B$)
could be related to the occupation number of different-statistic
particles of gases [10].
To examine the critical region possibly tuned by the
Pauli-blocking measure $B=\gamma N_0$ and the free orientation  $\theta$,
as evidenced from previous Boltzmann results [12] (cf. Chu therein) : $\lambda_i =0$ for
cases of   $\theta=\pi/4$ (or $B=-1$), we firstly check those
spectra near $\theta=0$, say, $\theta=0.005$ and $\theta=\pi/4
\approx 0.7854$, say, $\theta=0.78535$ for a $B$-sweep ($B$
decreases from 1 to -1),
respectively. Note that, as the  free-orientation
$\theta$ is not zero, there will be two kinds of propagation of
the disturbance wave : sound and diffusion modes [13-14]. The
latter (anomalous) mode has been reported in Boltzmann gases
(cf. Ref. 12 by Chu)
and is related to the propagation of entropy wave which is
not used in the acoustical analog here. The absence of (further)
diffusion (or maximum absorption) for the sound mode at certain
state ($h$, corresponding to the inverse of energy $E$; cf. Chu in Ref.
12) is classified as a localized state (resonance occurs) based
on the acoustical analog [12]. The state
of decreasing $h$ might, in one case [15], correspond to that of $T$ (absolute
temperature) decreasing as the mean free path is increasing
(density or pressure decreasing).
\newline We have observed the max. $\lambda_i$ (absorption of sound
mode, relevant to the localization length according to the
acoustical analog [12]) drop to around four orders of magnitude
from $\theta=0.005$ to $0.78535$ (please see Chu (2001) in [12])!
This is a clear demonstration of
the effect of free orientations. Meanwhile, once the Pauli-blocking measure
($B$) increases or decreases from zero (Boltzmann gases), the
latter (Fermi gases : $B <0$) shows opposite trend compared to
that of the former (Bose gases : $B>0$) considering the shift of
the max. $\lambda_i$ state : $\delta h$. $\delta h >0$ is  for
Fermi gases ($|B|$ increasing), and  the reverse ($\delta h <0$)
is for Bose gases ($B$ increasing)! This illustrates partly the
 interaction effect (through the Pauli exclusion
principle). These results will be crucial for further obtaining
the phase diagram (as the density or temperature is changed)
tuned by both the free orientation and
the  interaction. Here, $B=-1$ or
$\theta=0, \pi/4$ might be fixed points.
\newline
To check what happens
when the temperature is decreased (or  $h$ is
decreased) to near $T=0$ or $T=T_c$, we collect all the data based on the
acoustical analog from the dispersion relations (especially the
absorption of sound mode) we calculated for ranges in different
degrees of the orientation (here, $\theta$ is up to $\pi/4$ considering
single-particle scattering and binary collisions; in fact, effects
of $\theta$ are symmetric w.r.t. $\theta=\pi/4$ for
$0\le\theta\le\pi/2$; cf. Chu (2001) in Ref. 12) and
Pauli-blocking measure. After that,
we plot the possible phase diagram for the inverse of the
 rarefaction parameter
 vs. the  orientation (which is related to the scattering) into
Fig. 1 (for different $B$s : $B=-0.98, -0.9, -0.1, 0, 0.1, 1$).
Here, the Knudsen number ($K_n$) $\propto$ MFP/$\lambda_s$ with MFP and $\lambda_s$
 being
the mean free path and wave length, respectively
 and the temperature vs. MFP relations could be, ine one case,
traced  from Ref. 15 (following Fig. 3 therein).
This figure shows that as the temperature decreases to a
rather low value, the orientation will
decrease sharply (at least for either Bose or Fermi gases).
There is no doubt that this result resembles qualitatively that proposed
before [16] for cases of the pressure vs. temperature analogue.\newline
On the other hand, qualitatively similar  results (cf. Fig. 4 in [17] :
theirin higher relative pressure corresponding to $\lambda_r$ here)
show that (i) once the  orientation $\theta$ increases, for the same $h$
(or temperature [15]), the dispersion $\lambda_r$ (or the relative
pressure [17]) increases (please refer to Chu (2001) in [12]); (ii) as
$|B|$ ($B$: the Pauli-blocking parameter) increases, the
dispersion ($\lambda_r$) will reach the continuum or
hydrodynamical limit (larger $h$ or high temperature regime)
earlier. The phase
speed of the plane (sound) wave in Bose gases (even for small but
fixed $h$) increases more rapid than that of Fermi gases (w.r.t.
to the higher temperature conditions : larger $h$) as the
relevant parameter B increases. For all the rarefaction measure
($h$), perturbed plane waves propagate faster in Bose-particle gases than
Boltzmann-particle and Fermi-particle gases (e.g., see [18] or [13]). In fact, the real part ($\lambda_r$)
also resembles qualitatively  those reported in [19] for $T>T_c$ cases.  \newline
As for the imaginary part
($\lambda_i$), there is the
maximum absorption (or attenuation) for certain $h$ (the rarefaction
parameter)
which resembles that reported in [20] (cf. Fig. 1 (b) therein).  We observed a jump of the (relative)
sound speed
in the multiple scattering case (cf. Chu (2002) in [12])
which was also reported in [21](at the phase transition between hadronic
phase and QGP).
\newline
To know the detailed effects of  interactions
(tuned by the Pauli-blocking parameter : $B$ here) and the orientation,
which could be linked to the effective number of thermodynamic
degrees of freedom (as stated in [11] : $\nu$, for an ideal gas of
massless, non-interacting constituents, $\nu$ counts the number of
bosonic degrees of
freedom plus the number of fermionic degrees of freedom
weighted by $7/8$), we plot $\theta$ (of which the localization or
resonance occurs for specific $B$) vs. $T$ (the temperature, in arbitrary units)
in Figs. 2 and 3 by referring to two possibly localized states ($\theta=0$
or $\pi/4$; cf. Chu (2001) in [12]). The trends here resemble that of
 Fig. 1
 in [11] once  the effective number
 of thermodynamic
degrees of freedom   is proportional to the product of the
orientation and the Pauli-blocking parameter
($\nu\propto \,\theta \times |B|$).
Then there might be different behaviors separated by crossover regions
as shown in Fig. 2.  We remind the readers that for the case of $\theta=\pi/4$
dominated (Fig. 3), as the lower temperature is associated with the higher
density, fermions ($B<0$) link to the lower temperature regime
(under the same orientation).
\newline
It is clear that depending on the temperature regime investigated,
there would be different values of $\nu$. If there are differences between ours and those in [11]
(presumed that the matter under study is in (approximate) local thermal equilibrium.
At RHIC, such evidence is believed to be provided
by the agreement of the elliptic flow [22] measured in noncentral
collisions with hydrodynamic model predictions. Such predictions are based on the assumption that the
matter behaves like a fluid in local thermal equilibrium,
with arbitrarily short mean free paths and correspondingly
strong interactions.), one possible
explanation could be that the assumption of a completely
thermal medium is a simplification [23]. The acoustic perturbations
we treated are close to the thermodynamic equilibrium (for Bose or
Fermi gases). Other reasoning is related to the different types of
particles (with or without  fragmentation) being considered. One interesting observation
is that the attenuation of jet (quenching) observed at RHIC resembles qualitatively
the attenuation of plane (sound) waves (cf. Chu in [12]).\newline
With above results, then our approach could provide an effective theory based on the opposite picture of very
strong interaction (via the tuning of $B$ and $\theta$) and very small mean free paths ($h$ is large) [24].
This can also be useful to the study of problems in astrophysics : like compact stars. For instance,
one of the most striking features of QCD is asymptotic freedom: the force
between quarks becomes arbitrarily weak as the characteristic momentum
scale of their interaction grows larger. This immediately suggests that at
sufficiently high densities and low temperatures (corresponding to the case
of Fig. 3 here; cf. Chu (2001) in [12] since $\theta=\pi/4$ is also possible
and thus dominates the localized behavior or transition) matter will consist of a Fermi
sea of essentially free quarks, whose behavior is dominated by the freest of
them all: the high-momentum quarks that live at the Fermi surface.
If the strange matter hypothesis is true, a new class of compact stars called
strange
stars should exist.  They would
form a distinct and disconnected branch of compact stars and are not a part
of the
continuum of equilibrium configurations that include white dwarfs and
neutron stars. In principle, strange and neutron stars could coexist.
%
\newline To conclude in brief, our illustrations here, although
are based on the acoustical analog  of our quantum discrete
kinetic calculations, can indeed show the Fermi and Bose liquid
(say, Cooper pairs) and
their
 critical behavior for the  transition
(at least valid to the regime $T>T_c$ considering the QGP) once the temperature is
decreased to rather low values. We
shall investigate more complicated problems in the future [24-26] (e.g., the saturated orientation
shown in Fig. 3 which is almost the same for all different-statistic gases of particles might be
relevant to the Stefan-Boltzmann limit; cf. Fig. 5 in [25] by Rischke).

\newpage

\psfig{file=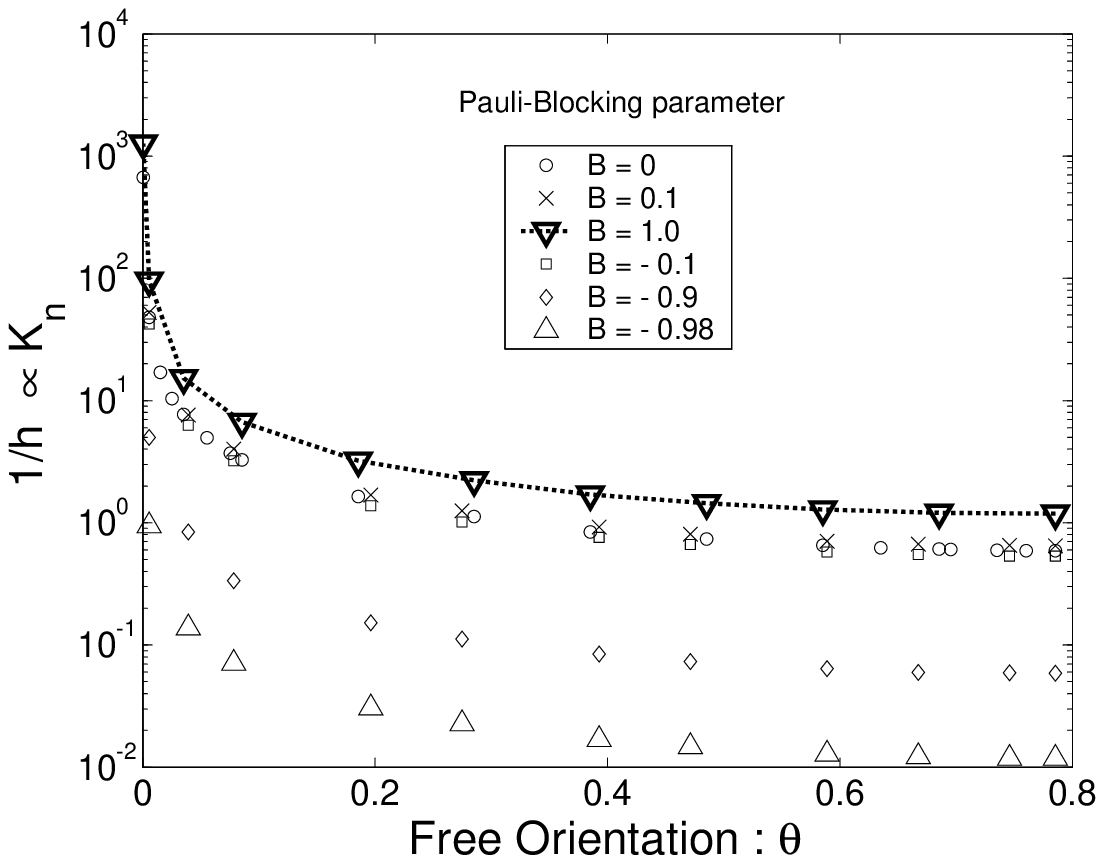,bbllx=0.1cm,bblly=12.9cm,bburx=12cm,bbury=23.8cm,rheight=9.2cm,rwidth=9.2cm,clip=}

\begin{figure}[h]
\hspace*{10mm} Fig. 1 \hspace*{1mm} Possible phase diagram for
different-statistic  gases w.r.t. the free \newline \hspace*{10mm}
orientation ($\theta$) and
Knudsen number ($K_n
\propto$ the mean free path/wave length). 
\newline \hspace*{10mm} $B >0$ : bosonic particles; $B<0$ : fermionic
 particles [10].  The orientation is \newline \hspace*{10mm} related to the scattering. $K_n$
might be transformed to the dimensionless or relative temperature (cf. Fig. 3 in [15]
).
\end{figure}

\newpage

\psfig{file=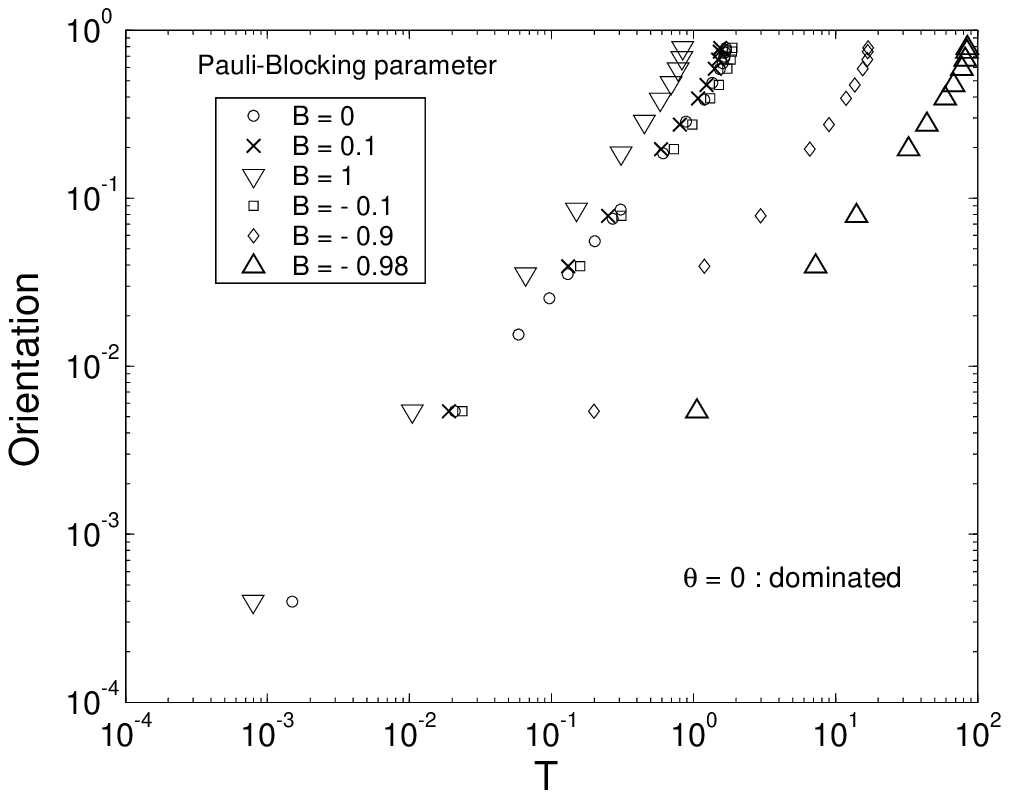,bbllx=0cm,bblly=13.9cm,bburx=12cm,bbury=24.8cm,rheight=9.2cm,rwidth=9.2cm,clip=}
\begin{figure}[h]
\hspace*{10mm} Fig. 2 \hspace*{1mm} Possible phase diagram for
different-statistic  gases w.r.t. the \newline \hspace*{10mm}
 orientation ($\theta$) and
temperature $T \propto h$ (the rarefaction measure). 
\newline \hspace*{10mm} The unit of $T$ is arbitrary. $B >0$ : bosonic particles; $B<0$ : fermionic
 particles [10].\newline \hspace*{10mm} The trend here resembles that of
 Fig. 1
 in [11] once $\nu$ (the effective number \newline \hspace*{10mm}
 of thermodynamic
degrees of freedom ) is $\propto \,\theta \times |B|$.
\end{figure}

\newpage

\psfig{file=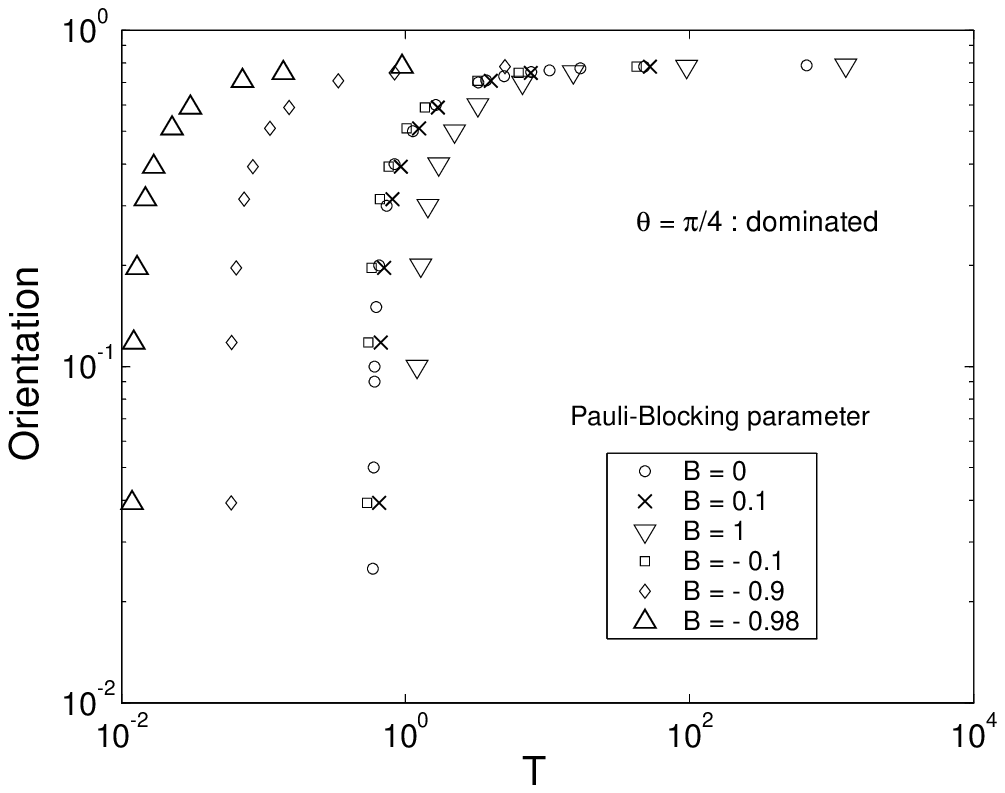,bbllx=0cm,bblly=13.9cm,bburx=12cm,bbury=24.8cm,rheight=9.2cm,rwidth=9.2cm,clip=}
%
\begin{figure}[h]
\hspace*{10mm} Fig. 3\hspace*{1mm} Possible phase diagram for
different-statistic  gases w.r.t. the \newline \hspace*{10mm}
 orientation and
temperature $T \propto K_n$ (the Knudsen number). 
\newline \hspace*{10mm} The unit of $T$ is arbitrary (cf. Fig. 3 in [15]).
\end{figure}


%

\begin{thebibliography}{99}    
\bibitem{Rev:N-Star} K. Kotake, K. Sato and K. Takahashi,
Rep. Prog. Phys. {\bf 69},  971 (2006).  
\bibitem{QC}  J.R. Oppenheimer and G.M. Volkoff, Phys. Rev.
{\bf 55}, 374 (1939).
R.C. Tolman, Phys. Rev. 55,  364 (1939).
\bibitem{N:Star} S.L. Shapiro and S.A. Teukolsky, {\it Black Holes,
White Dwarfs and Neutron Stars: The Physics
of Compact Objects} (Wiley,  New York, 1983).
F. Weber, 
{Prog. Particle Nucl. Phys.} {\bf 54},  193 (2005).  
H. Heiselberg and V. Pandharipande, Annu. Rev. Nucl. Part. Sci.  {\bf 50},
481 (2000).  
\bibitem{SC:QP}  M. Alford, Dense Quark Matter in Compact Stars,
in : W. Plessas and L. Mathelitsch (Eds.), Lect. Notes Phys. {\bf 583},
pp. 81-115, (Springer, Berlin, 2002).
\bibitem{Landau:1957} L.D. Landau, Sov. Phys. JETP {\bf 5}, 101 (1957).
C.M. Varma, Z. Nussinov, and W. van Saarloos, Phys. Rep. {\bf 361}, 267 (2002).
\bibitem{R:N-Star} M.A. McLaughlin, A. G. Lyne {\it et al.},
Nature {\bf 439}, 817 (2006).
\bibitem{Magn:P} S. Johnston and R.W. Romani,
Astrophys. J. {\bf 590}, L95 (2003).  
\bibitem{Pauli:B} M.S. Hussein, R.A. Rego and C.A. Bertulani,
Phys. Rep. {\bf 201},  279 (1991). M.A.G. Alvarez, N. Alamanos,
L.C. Chamon and M.S. Hussein, nucl-th/0501084.
\bibitem{BU:U}  A. Dobado and F.J. Llanes-Estrada, Phys. Rev.  D {\bf 69},  116004 (2004).
 D.T. da Silva and D. Hadjimichef, J. Phys. G-Nucl. Part. Phys. {\bf 30},  191 (2004).   
\bibitem{U:U} V.V. Vedenyapin, I.V. Mingalev  and  O.V. Mingalev,  Russian
Academy of Sciences Sbornik Mathematics {\bf 80} 271 (1995).
A. K.-H. Chu, Phys. Scr. {\bf 69}, 170 (2004).   
\bibitem{Therm:Entrop} B. M\"{u}ller and  K. Rajagopal,
Eur. Phys. J. C {\bf 43}, 15 (2005).  
\bibitem{RMP:Loc2001} J.D. Maynard,  Rev. Mod. Phys. {\bf 73},
401 (2001).
 K.-H. W. Chu, {J. Phys. A Math. and General} {\bf 35}, 1919 (2002).
 K.-H. W. Chu, {J. Phys. A Math. and General} {\bf 34}, L673 (2001).
\bibitem{Chu:KH} R. K.-H. Chu, Meccanica {\bf 39},  383 (2004).  
A. K.-H. Chu, cond-mat/0411627.
\bibitem{Diffus:S}R.J. Jr. Mason, in {\it Rarefied Gas Dynamics},
edited by J.H. de Leeuw (Academic Press, New York, 1965), vol. 1, p. 48.  
\bibitem{MFP:Slip}  D. Einzel and  J.M. Parpia,  { J. Low Temp. Phys.} {\bf
109}, 1 (1997).
\bibitem{Alf:HD} G. Boyd {\it et al.}, Phys. Rev. Lett. {\bf 75}, 4169 (1995).
M. Okamoto {\it et al.}, Phys. Rev. D {\bf 60}, 094510  (1999).
M. Alford, Nucl. Phys. B (Proc. Suppl.) {\bf 117}, 65 (2003).  
\bibitem{Lattice:QCD} F. Karsch, Nucl. Phys. A {\bf 698}, 199 (2002).
\bibitem{Chu:2004} K.-H. W.  Chu,  Preprint (2004).
\bibitem{Dispers:S} M. Asakawa and T. Hatsuda,
Nucl. Phys. A {\bf 610}, 470 (1996).  
\bibitem{Attenu:Exp} W. Broniowski and B. Hiller,
Phys. Lett. B {\bf 392}, 267 (1997).  
\bibitem{Jump:S} Yu.A. Tarasov,
Phys. Lett. B {\bf 379}, 279 (1996).  
\bibitem{Ellipt:Flow} P.F. Kolb, P. Huovinen, U. Heinz and H. Heiselberg,
Phys. Lett. B {\bf 500}, 232 (2001).  
X.-N. Wang, Nucl. Phys. A {\bf 750}, 98 (2005). 
\bibitem{Deconf:Hadron}  B. M\"{u}ller, Nucl. Phys. A {\bf 750}, 84 (2005). 
\bibitem{QGP:Phase} E. Shuryak,
Prog. Particle Nucl. Phys. {\bf 53},  273  (2004).  
E. Shuryak,
Nucl. Phys. A {\bf 750}, 64 (2005).  
\bibitem{QGP:Equil} D.H. Rischke,
Prog. Particle Nucl. Phys. {\bf 52}, 197 (2004).  
T.K. Nayak,
Pramana-J. Phys. {\bf 62},  623 (2004).  
\bibitem{Q:B}  J. Liao and E.V. Shuryak,
Phys. Rev. D {\bf 73}, 014509 (2006).
M. Stephanov,
Acta Physica Polonica B
{\bf 35}, 2939 (2004).  
M. Gyulassy and L. McLerran,
Nucl. Phys. A {\bf 750}, 30 (2005).  
M. Bluhm, B. K\"{a}mpfer and  G. Soff,
Phys. Lett. B {\bf 620}, 131 (2005).  
\end{thebibliography}
\end{document}